\documentstyle[preprint,aps]{revtex}

\begin{document}
\draft         % PRINTS PACS NUMBER
\preprint{IFUSP-P 1131 and IFT-P.017/95 (hep-ph/9503432)}
\title{Quartic Anomalous Couplings in \boldmath{$\gamma\gamma$} Colliders}
\title{Quartic Anomalous Couplings in \boldmath{$\gamma\gamma$} Colliders}
\author{O.\ J.\ P.\ \'Eboli \cite{ojpe},  M.\ B.\ Magro \cite{mbm},
P.\ G.\ Mercadante \cite{pgm}, \\
 }
\address{Instituto de F\'{\i}sica,
Universidade de S\~ao Paulo, \\
C.P. 20516,  01452-990 S\~ao Paulo, Brazil}
\author{S.\ F.\ Novaes \cite{sfn} }
\address{Instituto de F\'{\i}sica Te\'orica,
Universidade  Estadual Paulista, \\
Rua Pamplona 145,  01405-900 S\~ao Paulo, Brazil}

\date{\today}
\maketitle

\begin{abstract}

We study the constraints on the vertices $W^+W^- Z\gamma$,
$W^+W^-\gamma\gamma$, and $ZZ\gamma\gamma$ that can be obtained
from triple-gauge-boson production at the next generation of
linear $e^+e^-$ colliders operating in the $\gamma\gamma$ mode.
We analyze the processes $\gamma\gamma \rightarrow W^+W^-V$
($V=Z$, or $\gamma$) and show that these reactions increase the
potential of $e^+e^-$ machines to search for anomalous
four-gauge-boson interactions.

\end{abstract}

\vskip 1.in

\begin{center}
{\em Submitted to Physical Review D1}
\end{center}

%\pacs{XXX}

\newpage

\section{Introduction}
\label{sec:int}

In the next generation of $e^+ e^-$ colliders, multiple vector-boson
production will provide a crucial test of the gauge structure of the
Standard Model (SM), since it will allow the study of the triple and
quartic vector-boson couplings. These vertices are strictly
constrained by the $SU(2)_L \otimes U(1)_Y$ gauge invariance and any
small deviation from the SM predictions spoils the precise
cancellation of the high-energy behavior between the various
diagrams, giving rise to an anomalous growth of the cross section with
energy. Therefore, the careful study of multiple vector-boson
production, and consequently of vector-boson self-interactions, can
give important clues about the existence of new particles and/or
interactions beyond the SM.

The reaction $e^+ e^- \rightarrow W^+ W^-$ will be accessible at
LEP II and some information about the $WW\gamma$ and $WWZ$
vertices will be available in the near future \cite{ano:ee}.
Nevertheless, we will have to wait for colliders with higher
center--of--mass  energies in order to produce a final state with
three or more gauge bosons and to test the quartic gauge--boson
interaction. This will be accomplished at the Next Linear
$e^+e^-$ Collider (NLC) \cite{pal}, which will reach an energy
between 500 and 2000 GeV with an yearly integrated luminosity of
at least $10$ fb$^{-1}$. An interesting feature of these new
machines is the possibility of transforming an electron beam into
a photon one through the laser backscattering mechanism
\cite{las0,laser}.  This process will allow the NLC to operate in
three different modes, $e^+e^-$, $e\gamma$, and $\gamma\gamma$,
opening up the opportunity for a wider search for new physics.
However, it is important to point out that the collider can
operate in only one of its three modes at a given time,
therefore, it is essential to study comparatively the different
features of each of these setups.

The triple and quartic gauge--boson vertices, in the framework of gauge
theories, have a common origin, and consequently, a universal strength
dictated by the gauge symmetry of the model.  Notwithstanding, in a
more general context, anomalous quartic couplings can arise as the
low--energy limit of heavy state exchange, whereas trilinear couplings
are modified by integrating out heavy fields. In this sense, it is
possible to conceive extensions of the SM where the trilinear
couplings remain unchanged, while the quartic vertices receive new
contributions.  For instance, the introduction of a new heavy scalar
singlet, that interacts strongly with the Higgs sector of the SM,
enhances the four vector-boson interaction without affecting either
the triple vector-boson couplings or the SM predictions for the $\rho$
parameter \cite{singlet}. Therefore, the measurement of the
three--vector--boson production cross section can provide further
non--trivial tests of the SM that are complementary to the analysis of
the production of vector--boson pairs.

The cross section for the production of multiple gauge bosons, in
the context of the SM, has already been evaluated for $e^+e^-$
colliders operating in the $e^+ e^-$ \cite{bar:plb,bar:num},
$e\gamma$ \cite{cheung}, and $\gamma\gamma$ \cite{our:gg} modes.
There has also been some studies of anomalous quartic vertices
through the reactions $e^+ e^- \rightarrow VVV$
\cite{bela1,stir}, $e\gamma \rightarrow V V F$ \cite{our:eg}, and
$\gamma\gamma \rightarrow VV$ \cite{bela2}, where $V=$ $Z$,
$W^\pm$ or $\gamma$ and $F=$ $e$ or $\nu_e$.  In this work, we
analyze the effect of some genuinely anomalous quartic operators,
{\it i.e.} operators which do not modify the trilinear vertices
and cannot be bounded by the LEP II measurements. Furthermore,
since we are interested in probing these anomalous couplings in a
$\gamma \gamma$ collider, we concentrate our analyses on
operators that involve at least one photon.  We studied the
production of three vector bosons in $\gamma\gamma$ collisions
through the reactions
\begin{equation}
\eqnum{I}
\label{g}
\gamma + \gamma \rightarrow W^+  +  W^- + \gamma \; ,
\end{equation}
\begin{equation}
\eqnum{II}
\label{z}
\gamma + \gamma \rightarrow W^+  +  W^-  + Z \; ,
\end{equation}
in order to impose bounds on the vertices $\gamma Z W^+ W^-$,
$\gamma\gamma W^+W^-$, and $\gamma\gamma Z Z$. These processes
involve only interactions between the gauge bosons,  making more
evident any deviation from  the SM predictions. Moreover, there
is no tree--level contribution involving the Higgs boson which
evades all the uncertainties coming from the scalar sector, like
the Higgs boson mass.

Our results show that the constraint on the anomalous couplings
$\gamma Z W^+ W^-$, coming from the reaction \ref{z}, is as
restrictive as the one obtained in $e\gamma$ colliders
\cite{our:eg} and a factor of 5 better than the one coming from
$e^+e^-$ colliders \cite{bela1,stir}. On the other hand, the
bounds in the vertices $\gamma\gamma W^+W^-$ and $\gamma\gamma Z
Z$, stemming from the processes \ref{g} and \ref{z}, are a factor
from 2 to 5 better than the ones that can be obtained in an $e^+
e^- $ collider, however, they are worse than the ones coming from
the direct reaction $\gamma\gamma \rightarrow W^+ W^-$
\cite{bela2}.

The next section contains the effective operators that we
analyzed and the calculational method employed in this paper.  Our
results and discussions are presented in Sec.\ III.

\section{Effective Lagrangians and Calculational Method}

In order to construct effective operators associated to
exclusively quartic anomalous couplings we employed the formalism
of Ref.\ \cite{schild}. We required the existence of a custodial
$SU(2)_{C}$ symmetry, which avoids any contribution to the $\rho$
parameter, and we also demanded that the phenomenological
Lagrangians are invariant under local $U(1)_{\text em}$ symmetry.
The lowest order operators that comply with the above
requirements, involving at least one photon, are of dimension six
\cite{boud}: there are two independent $C$ and $P$ conserving
operators involving two photons
\cite{bela2}
\begin{eqnarray}
{\cal L}_0 & = & - \frac{\pi \alpha}{4 \Lambda^2}\,
a_0\, F^{\mu\nu}F_{\mu\nu}  W^{(i)\alpha} W^{(i)}_\alpha  \; ,
\label{lag:0} \\
{\cal L}_c & = & - \frac{\pi \alpha}{4 \Lambda^2}\,
a_c\, F^{\mu\alpha}  F_{\mu\beta} W^{(i)}_\alpha W^{(i)\beta} \; ,
\label{lag:c}
\end{eqnarray}
and one operator exhibiting just one photon \cite{our:eg}
\begin{equation}
{\cal L}_n  = \frac{\pi \alpha}{4 \Lambda^2}\, a_n\,
 \epsilon_{ijk} W^{(i)}_{\mu\alpha} W^{(j)}_\nu W^{(k)\alpha} F^{\mu\nu}
\; ,
\label{lag:new}
\end{equation}
where $W^{(i)}$ is the $SU(2)_{C}$ triplet and $F^{\mu\nu}$ is
the electromagnetic field strength.  In terms of the physical
fields $W^+$, $W^-$, and $Z$, the effective Lagrangians
(\ref{lag:0}) and (\ref{lag:c}) give rise to anomalous
$W^+W^-\gamma\gamma$ and $ZZ\gamma\gamma$ couplings, while
(\ref{lag:new}) generates a new $W^+W^-Z\gamma$ vertex. We should
notice that the $ZZ\gamma\gamma$ vertex is particularly
interesting since it is completely absent in the SM. Moreover, we
note that the effective operators (\ref{lag:0}) and
(\ref{lag:new}) can parametrize the exchange of a neutral scalar
particle while the operator (\ref{lag:c}) corresponds to the
exchange of a charged particle. Therefore, we expect that the
characteristic scale of the above interactions is set by the
masses of the exchanged states.  In the limit of very heavy scalars,
the expected values of the coefficients $a_i$ ($i=0$, $c$, $n$)
should be in the range of $10^{-2}$ to $10^{-3}$.  For the sake
of definiteness, our results are presented assuming  that
$\Lambda = M_W$.

An interesting feature of the above effective Lagrangians is that they
will not be directly constrained by LEP II since they do not
contribute to triple gauge--boson vertices. In order to obtain some
bounds on these couplings at low energies, we must rely on their
contribution to one-loop processes. From the analysis of oblique
radiative corrections to the $Z$ physics \cite{our:eg}, we estimated
that $-4.5 < a_0 < 0.64$ and $-11 < a_c < 5.8$ at $1\sigma$ level,
while the coupling $a_n$ is not constrained at all.

In order to evaluate the triple--vector--boson production at NLC, we
assumed that this machine will reach a center--of--mass energy of
$500$ ($1000$) GeV with a yearly integrated luminosity of 10
fb$^{-1}$. The most promising mechanism to generate hard photon beams
in an $e^+ e^-$ linear collider is laser backscattering.  Assuming
unpolarized electron and laser beams,  the backscattered photon
distribution function \cite{laser} is
\begin{equation}
F_{\gamma/e}  (x,\xi) \equiv \frac{1}{\sigma_c} \frac{d\sigma_c}{dx} =
\frac{1}{D(\xi)} \left[ 1 - x + \frac{1}{1-x} - \frac{4x}{\xi (1-x)} +
\frac{4
x^2}{\xi^2 (1-x)^2}  \right] \; ,
\label{f:l}
\end{equation}
with
\begin{equation}
D(\xi) = \left(1 - \frac{4}{\xi} - \frac{8}{\xi^2}  \right) \ln (1 + \xi) +
\frac{1}{2} + \frac{8}{\xi} - \frac{1}{2(1 + \xi)^2} \; ,
\end{equation}
where $\sigma_c$ is the Compton cross section, $\xi \simeq 4
E\omega_0/m_e^2$, $m_e$ and $E$ are the electron mass and energy
respectively, and $\omega_0$ is the laser-photon energy. The quantity
$x$ represents the ratio between the scattered photon and initial
electron energy and its maximum value is
\begin{equation}
x_{\text{max}}= \frac{\xi}{1+\xi} \; .
\end{equation}
In what follows, we assumed that the laser frequency is such that $\xi
= 2(1 +\sqrt{2})$, which leads to the hardest possible spectrum of
photons with a large luminosity.

The cross section for the triple--vector--boson production via
$\gamma\gamma$ fusion can be  obtained by folding the elementary
cross section  for the subprocesses $\gamma\gamma \rightarrow
WWV$ ($V=  \gamma, \; Z$) with the photon-photon luminosity
($dL_{\gamma\gamma}/dz$), {\it i.e.},
\begin{equation}
d\sigma (e^+e^-\rightarrow \gamma\gamma \rightarrow WWV)(s) =
\int_{z_{\text{min}}}^{z_{\text{max}}} dz ~ \frac{dL_{\gamma\gamma}}{dz} ~
d \hat\sigma  (\gamma\gamma \rightarrow WWV) (\hat s=z^2 s) \; ,
\end{equation}
where $\sqrt{s}$ ($\sqrt{\hat{s}}$) is the $e^+e^-$ ($\gamma\gamma$)
center-of-mass energy, $z^2= \tau \equiv \hat{s}/s$, and the
photon-photon luminosity is
\begin{equation}
\frac{d L_{\gamma\gamma}}{dz} = 2 ~ z  ~
\int_{z^2/x_{\text{max}}}^{x_{\text{max}}} \frac{dx}{x}
F_{\gamma/e} (x,\xi)F_{\gamma/e} (z^2/x,\xi) \; .
\label{lum}
\end{equation}

The analytical calculation of the cross section for the
subprocess $\gamma\gamma \rightarrow W^+W^- \gamma$
($\gamma\gamma \rightarrow W^+W^-Z$) is very lengthy and
tedious despite being straightforward. In order to perform these
calculations in an efficient and reliable way, we evaluated
numerically the helicity amplitudes using the techniques outlined
in Ref.\ \cite{bar:num,zep:num}. The phase space integrations
were performed numerically using the Monte Carlo routine VEGAS
\cite{lepage}. As a check of our results,  we explicitly verified
that the amplitudes were Lorentz and $U(1)_{\text em}$ invariant
and that the kinematic distributions for $W^+$ and $W^-$
coincided.

\section{Results and Discussion}

The total cross sections for the processes \ref{g} and \ref{z}
are quadratic functions of the anomalous couplings $a_i$, {\it
i.e.}
\begin{equation}
\sigma_{\text{tot}} = \sigma_{\text{sm}} + a_i \; \sigma^i_{\text{int}}
+  a^2_i \; \sigma^i_{\text{ano}} \; ,
\label{base}
\end{equation}
where $\sigma_{\text{sm}}$ stands for the SM cross section
\cite{our:gg} and $\sigma^i_{\text{int}}$
($\sigma^i_{\text{ano}}$) is the interference (pure anomalous)
contribution. We evaluated the cross sections involved in Eq.\
(\ref{base}), imposing that the polar angles of the produced
vector bosons with the beam pipe are larger than $10^\circ$  and
assuming $\Lambda =  M_W$.  For the process \ref{g}, we also
introduced a cut on the photon transverse momentum, $p_T^\gamma >
10$ ($20$) GeV, not only to guarantee that our results are free
of infrared divergences but also to mimic the performance of a
typical electromagnetic calorimeter.  In Table \ref{sigmas:z}, we
present our results assuming in each case that only one anomalous
coupling is nonvanishing.

In order to quantify the effect of the new couplings, we defined
the  statistical significance $S$ of the anomalous signal
\begin{equation}
S = \frac{|\sigma_{\text tot} -
\sigma_{\text sm}|}{\sqrt{\sigma_{\text sm}}} \;
\sqrt{\cal L} \; ,
\label{sig}
\end{equation}
which can be easily evaluated using the parametrization (\ref{base})
with the coefficients given in Table \ref{sigmas:z}.  We list in Table
\ref{const} the values of the anomalous couplings that correspond to a
$3\sigma$ effect in the total cross section for the different
processes, assuming an integrated luminosity ${\cal L}= 10$ fb$^{-1}$
for the correspondent $e^+e^-$ collider. It is interesting to point
out that the most restringent bounds come from the process \ref{g} and
that our results are quite insensitive to different choices of the
angular and $p_T^\gamma$ cuts, as we can learn from Table \ref{const}.

We can see from Table \ref{const} that our $3\sigma$ limits for
the coupling $a_n$ are a factor of $3$ better than the limits
obtained in $e^+e^-$ collisions \cite{bela1,stir}, while they are
of the same order of the ones originating from an $e\gamma$
machine \cite{our:eg}. The $\gamma\gamma$ mode of the NLC is more
efficient for studying this anomalous quartic coupling since it
leads to cross sections for the production of three vector
bosons that is more than one order of magnitude larger than the
ones for a conventional $e^+e^-$ collider with $\sqrt{s} \geq 500$
\cite{our:gg}.

The limits on the anomalous coupling $a_c$ that can be
established through the triple gauge--boson production are one
order of magnitude better than the ones coming from the $e^+ e^-$
mode and are comparable to the constraints that can be obtained in
the $e\gamma$ mode. However, the bounds shown in Table
\ref{const} are a factor of 2 weaker than the ones obtained from
the direct reaction $\gamma\gamma \rightarrow W^+ W^-$ ($ZZ$),
since the triple gauge--boson production is a higher order
process in $\alpha_{\text{em}}$. Our limits on the coupling $a_0$
are slightly better than the ones arising from the $e^+ e^-$
mode, while they are almost one order of magnitude worse than
the ones obtained in $e\gamma$ \cite{our:eg} or direct
$\gamma\gamma \rightarrow W^+W^-$ ($ZZ$) \cite{bela2} collisions.

The kinematical distributions of the final state particles can be
used, at least in principle, to increase the sensitivity of the
$\gamma\gamma$ reactions to the anomalous couplings, improving
consequently the bounds on the new interactions. More than that, they
could furnish further information that would allow us to distinguish
among the different anomalous interactions. In Figs.\ \ref{fig:1} to
\ref{fig:4}, we exhibit some representative distributions for the
processes \ref{g} and \ref{z}, adopting the values of the anomalous
coupling constants that lead to a $3\sigma$ deviation in the total
cross section.

We present in Fig.\ \ref{fig:1} the normalized $\cos
\theta_{W^\pm} $ distribution for the production of
$W^+W^-\gamma$, where $\theta_{W^\pm}$ is polar angle of the
vector boson with respect to the beam pipe. From this figure, we
can see that the anomalous $W$ distributions differ from the SM
one. For instance, we can identify  the cases with negative
values of $a_0$ and $a_c$, but we cannot discriminate between the
positive values of these anomalous couplings. Moreover, it is
interesting to notice that the anomalous couplings enhance the
production of $W^\pm$ in the central region of the detector,
where they can be more easily reconstructed.

Figure \ref{fig:2} contains the normalized distribution for the
invariant mass ($M$) of vector--boson pairs $W^+W^-$ for the process
\ref{g}. Here we can observe that the presence of the anomalous
interactions increases the invariant mass of the $W^+W^-$ pairs since
the new couplings are proportional to the photon momentum. Furthermore,
this distribution allows us to separate the negative $a_0$ and $a_c$
couplings, while the positive values of the anomalous couplings lead
to distributions similar to the SM ones. We notice in our analyses
that the distributions involving photons are less sensitive to the
anomalous couplings than the $W^\pm$ ones.

Although the process $\gamma \gamma \to W^+W^-Z$ has a lower
sensitivity, when compared with the $W^+W^-\gamma$ production, it
exhibits some interesting features, as can be seen from Figs.\
\ref{fig:3} and \ref{fig:4}. We show in Fig.\ \ref{fig:3} the
normalized $Z$ transverse momentum distributions for the process
\ref{z} using the $a_0$, $a_c$, and $a_n$ anomalous couplings. We
can learn from this figure that the anomalous couplings favor the
$Z$'s to have a higher $p_T$ since the momentum dependence of the
anomalous vertices enhances their contribution at higher
energies. Once again, we can distinguish between the different
anomalous interactions provided the couplings $a_0$ and
$a_c$ are negative, since these couplings lead to larger purely
anomalous contributions. It is interesting to remark that our
analyses show that the $p_T^{W^\pm}$ distribution does not allow
us to separate clearly the different anomalous interactions.

In Fig.\ \ref{fig:4}, we show the $W^\pm$ rapidity distributions
for the process \ref{z}.  The anomalous interactions lead to a
more copious production of $W^\pm$ in the central region and it
is also possible to have an indication of which effective
interaction is responsible for the departure from the SM
predictions. We found out that, as happened in $W^+W^-\gamma$
process, it is not possible to distinguish between the anomalous
interactions through the rapidity distribution of the neutral
vector bosons.

The bottom line of this work is that the study of the production
of $W^+W^-\gamma$ and $W^+W^-Z$ in a $\gamma\gamma$ collider will
be able to increase the potential of $e^+e^-$ machines to search
for anomalous four--gauge--boson interactions. Furthermore, the
kinematical distributions of the final state particle can also
help to distinguish between some of these interactions.

%**********
\acknowledgments
This work was partially supported by Conselho Nacional de
Desenvolvimento Cient\'{\i}fico e Tecnol\'ogico (CNPq), and
by  Funda\c{c}\~ao de Amparo \`a Pesquisa do Estado de S\~ao
Paulo (FAPESP).

%%%%%%%%%%%%%%%%%%%%%
\appendix
\section*{}

We collect in this appendix the expressions for anomalous
contributions to the the amplitudes of the processes
$\gamma\gamma \rightarrow W^+W^-V$, with $V=Z$ or $\gamma$.  The
standard model expressions can be found in Ref.\ \cite{our:gg}.
The momenta and polarizations of the initial photons were denoted
by ($k_1$, $k_2$) and [$\epsilon_\mu(k_1)$, $\epsilon_\nu(k_2)$],
while the momenta and polarizations of the final state $W^+$,
$W^-$ and $V$ are given by ($p_+$, $p_-$, $k_3$) and
[$\epsilon_\alpha(p_+)$, $\epsilon_\beta(p_-)$,
$\epsilon_\gamma(k_3)$] respectively. The amplitude of these
processes can be written as
\begin {equation}
{\cal M}_i= G_{V}
\epsilon_{\mu}(k_1)\epsilon_{\nu}(k_2)\epsilon_{\alpha} (p_+)
\epsilon_{\beta}(p_-)\epsilon_{\gamma}(k_3)
\left[ M_{\text{sm}}^{\mu\nu\alpha\beta\gamma}
+ M_{\text{ano}}^{\mu\nu\alpha\beta\gamma} (i) \right]\;,
\end {equation}
where $M_{\text{sm}}$ is the standard model invariant amplitude,
and $M_{\text{ano}} (i), \; i= a_0, \; a_c, \; a_n$,  represents
the different anomalous contributions.  The factor $G_V$ depends upon the
process under consideration: $G_V = e^3$ for the production of
$W^+W^-\gamma$ and  $G_V = e^3 \cot^2{\theta_W}$ for the
production of   $W^+W^-Z$.

We can write a compact expression for the anomalous
amplitudes of the process (\ref{g}), in the form
\begin{eqnarray}
\left(M_{\text{ano}}^\gamma\right)^{\mu\nu\alpha\beta\gamma}
(a_0) = \left[ M_1(a_0) + M_2(a_0) + M_3(a_0) + M_4(a_0)
\right]^{\mu\nu\alpha\beta\gamma}\;,
\nonumber \\
\left(M_{\text{ano}}^\gamma \right)^{\mu\nu\alpha\beta\gamma}
(a_c) =  \left[ M_1(a_c) + M_2(a_c) + M_3(a_c) +
M_4(a_c)\right]^{\mu\nu\alpha\beta\gamma}\;,
\label{ano:g}
\end{eqnarray}
while for the process (\ref{z}) we have
\begin{eqnarray}
\left(M_{\text{ano}}^Z \right)^{\mu\nu\alpha\beta\gamma}(a_0) =
\left[M_3(a_0) + M_4(a_0) + M_5(a_0)\right]^{\mu\nu\alpha\beta\gamma}\;,
\nonumber \\
\left(M_{\text{ano}}^Z  \right)^{\mu\nu\alpha\beta\gamma}(a_c) =
\left[M_3(a_c) + M_4(a_c) + M_5(a_c)\right]^{\mu\nu\alpha\beta\gamma}\;,
\nonumber \\
\left(M_{\text{ano}}^Z \right)^{\mu\nu\alpha\beta\gamma} (a_n) =
\left[M_1(a_n) + M_2(a_n) + M_6(a_n)\right]^{\mu\nu\alpha\beta\gamma}\;,
\label{ano:z}
\end{eqnarray}
with
\begin{eqnarray}
M_1^{\mu\nu\alpha\beta\gamma} (i) &=& \Delta^{\beta\nu\xi}(-p_-,k_2)
D^W_{\xi\lambda}(k_2-p_-) \Gamma^{\lambda\alpha\mu\gamma}_{(i)}
(k_1,-k_3,k_2-p_-,-p_+)
+\; [ k_{1\leftrightarrow 2}\; ; \;\mu\leftrightarrow\nu ] \;,
\nonumber \\
M_2^{\mu\nu\alpha\beta\gamma} (i) &=& \Delta^{\mu\alpha\xi}(k_1,-p_+)
D^W_{\xi\lambda}(k_1-p_+) \Gamma^{\lambda\beta\nu\gamma}_{(i)}
(k_2,-k_3,-p_-,k_1-p_+)
+\; [ k_{1\leftrightarrow 2}\; ; \;\mu\leftrightarrow\nu ] \;,
\nonumber \\
M_3^{\mu\nu\alpha\beta\gamma} (i) &=&
\Delta^{\alpha\gamma\xi}(p_+,k_3)
D^W_{\xi\lambda}(p_++k_3)\Gamma^{\lambda\beta\nu\mu}_{(i)}(k_1,k_2,0,0) \;,
\nonumber \\
M_4^{\mu\nu\alpha\beta\gamma} (i) &=&
\Delta^{\gamma\beta\xi}(k_3,p_-)
D^W_{\xi\lambda}(-p_--k_3)\Gamma^{\lambda\alpha\nu\mu}_{(i)}(k_1,k_2,0,0) \;,
\nonumber \\
M_5^{\mu\nu\alpha\beta\gamma} (i) &=&
\frac{1}{\cos{\theta_W}^2}\Delta^{\alpha\beta\xi}(p_-,p_+)
D^Z_{\xi\lambda}(p_++p_-)\Gamma^{\lambda\gamma\nu\mu}_{(i)}(k_1,k_2,0,0) \;,
\nonumber \\
M_6^{\mu\nu\alpha\beta\gamma} (i) &=&
-\frac{a_n\sin\theta_W}{16 \cos{\theta_W}^2 \Lambda^2}
\biggl(  2 p_2^\mu g^{\alpha\beta}  g^{\nu\gamma}
- 2 p_2^\gamma  g^{\alpha\beta}  g^{\mu\nu}
-  p_2^\mu  g^{\alpha\gamma}  g^{\nu\beta}
+ p_2^\beta  g^{\mu\nu}  g^{\alpha\gamma}
\nonumber \\
- p_2^\mu  g^{\beta\gamma}  g^{\alpha\nu}
&+& p_2^\alpha g^{\mu\nu}  g^{\beta\gamma}
 - p_2^\alpha g^{\mu\beta}  g^{\nu\gamma}
+ p_2^\gamma g^{\mu\beta}  g^{\nu\alpha}
- p_2^\beta g^{\mu\alpha}  g^{\nu\gamma}
+ p_2^\gamma  g^{\mu\alpha}  g^{\nu\beta}
+ 1 \leftrightarrow 2 \biggr) \; .
\label{m123}
\end {eqnarray}

In Eq.\ (\ref{m123}), $D^V_{\xi\lambda}$, $V=W,Z$, represents the
vector boson propagator in the unitary gauge, $\Delta$ is the
usual triple--gauge--boson vertex function
\begin {equation}
\Delta^{\alpha\beta\gamma}(q_1,q_2)=i\left[(2q_1+q_2)^{\beta}g^{\alpha\gamma}
-(2q_2+q_1)^{\alpha}g^{\beta\gamma}+(q_2-q_1)^{\gamma}
g^{\beta\alpha}\right] \; ,
\end {equation}
and, $\Gamma_{(i)}$ are the  different four--gauge--boson
couplings for the anomalous interactions
\begin {eqnarray}
\Gamma^{\mu\nu\alpha\beta}_{(a_0)}(q_1,q_2,q_3,q_4)
=i \frac{a_0}{2\Lambda^2}  ~  g^{\mu\nu} ~
\left[ g^{\alpha\beta} ~ (q_1.q_2) - q_{2}^{\alpha} q_{1}^{\beta}\right] \; ,
\label{rf:a0}
\end {eqnarray}
\begin{eqnarray}
\Gamma^{\mu\nu\alpha\beta}_{(a_c)}(q_1,q_2,q_3,q_4)=
i \frac{a_c}{8 \Lambda^2} ~ \biggl[ &&
(q_1.q_2) \left(g^{\mu\alpha} ~  g^{\nu\beta} +
g^{\mu\beta} ~ g^{\alpha\nu}\right)
+ g^{\alpha\beta} ~  \left( q_{1}^{  \mu} q_{2}^{  \nu} +
q_{2}^{  \mu} q_{1}^{  \nu} \right)
\nonumber \\
&& - q_{1}^{  \beta} \left( g^{\alpha\mu} ~ q_{2}^{  \nu} +
g^{\alpha\nu} ~ q_{2}^{  \mu}  \right)
- q_{2}^{  \alpha} \left( g^{\beta\mu} ~ q_{1}^{  \nu} +
g^{\beta\nu} ~  q_{1}^{  \mu} \right)
\biggr] \; ,
\label{rf:ac}
\end {eqnarray}
\begin{eqnarray}
\Gamma^{\mu\nu\alpha\beta}_{(a_n)}(q_1,q_2,q_3,q_4)=
 \frac{a_n\sin\theta_W}{16 \cos{\theta_W}^2 \Lambda^2} ~ &\biggl\{ &
g^{\mu\beta} ~ \left[ g^{\nu\alpha} ~ q_1.(q_2 - q_3) -
q_{1}^{  \nu} (q_2 - q_3)^\alpha \right]
\nonumber \\
&-& g^{\nu\beta} ~ \left[ g^{\mu\alpha} ~  q_1.(q_2 - q_4) -
q_{1}^{  \mu} (q_2 - q_4)^\alpha  \right]
\nonumber \\
&+& g^{\mu\nu} ~ \left[ g^{\alpha\beta} ~ q_1.(q_3 - q_4 ) -
( q_3 - q_4 )^\alpha q_{1}^{  \beta} \right]
\nonumber \\
&-& q_{2}^{  \mu} \left(g^{\alpha\nu} ~ q_{1}^{  \beta} -
g^{\alpha\beta} ~ q_{1}^{  \nu}   \right)
+ q_{2}^{  \nu} \left(g^{\alpha\mu} ~ q_{1}^{  \beta} -
g^{\alpha\beta} ~ q_{1}^{  \mu}   \right)
\nonumber \\
&-& q_4^{  \beta} \left(g^{\alpha\mu} ~ q_{1}^{  \nu} -
g^{\alpha\nu}  ~ q_{1}^{  \mu}   \right)
+  q_3^{  \beta} \left(g^{\alpha\nu} ~ q_{1}^{  \mu} -
g^{\alpha\mu} ~  q_{1}^{  \nu}   \right)
\nonumber \\
&-&  q_3^{  \nu} \left(g^{\alpha\beta} ~ q_{1}^{  \mu} -
g^{\alpha\mu} ~ q_{1}^{  \beta}   \right)
+ q_4^{  \mu} \left(g^{\alpha\beta} ~ q_{1}^{  \nu} -
g^{\alpha\nu} ~ q_{1}^{  \beta}  \right)
\biggr\}  \; .
\label{rf:an}
\end{eqnarray}

The amplitude $M_5$ comes from the new anomalous vertex
$\gamma\gamma ZZ$, and ${\cal L}_n$ gives rise to a five--vertex
$\gamma\gamma W^+W^-Z$ represented by $M_6$, which is necessary
to preserve the gauge invariance of the amplitude.

%**********
% REFERENCES
%**********

%**********
% FIGURES
%**********

\protect
\begin{figure}
\protect
\caption{
Normalized angular distributions of the $W^\pm$ bosons with the beam
pipe (process \protect\ref{g}) for $\protect\sqrt{s}=500$ GeV
and $p_T^\gamma > 10$ GeV. The dotted line is the SM result and
the dashed line stands for both $a_0=0.062$ and $a_c=0.17$, while
the solid (dot-dashed) line represents the results for
$a_0=-0.17$ ($a_c= -0.83$).  }
\label{fig:1}
\end{figure}

%**********
\begin{figure}
\protect
\caption{
Normalized invariant mass distribution for the $W^+W^-$ pairs
(process \protect\ref{g}) for $\protect\sqrt{s}=500$ GeV and
$p_T^\gamma > 10$ GeV.  The dotted line is the SM result, and the
dashed (solid) line represents the results for $a_0=-0.17$ ($a_c=
-0.83$). The results for $a_0=0.062$ and $a_c=0.17$ cannot be
distinguished from the SM one.}
\label{fig:2}
\end{figure}

%**********
\begin{figure}
\protect
\caption{
Normalized transverse momentum distribution of the $Z$ boson
(process \protect\ref{z}) at $\protect\sqrt{s}=500$ GeV. The
histogram is the SM result, the dashed line stands for
$a_0 = 0.11$ and $a_c = 0.32$, the dotted (dot-dashed) line
represents the results for $a_0 = -0.34$ $(a_c = -1.4)$, and the
solid line stands for $a_n = 1.55$.}
\label{fig:3}
\end{figure}

%**********
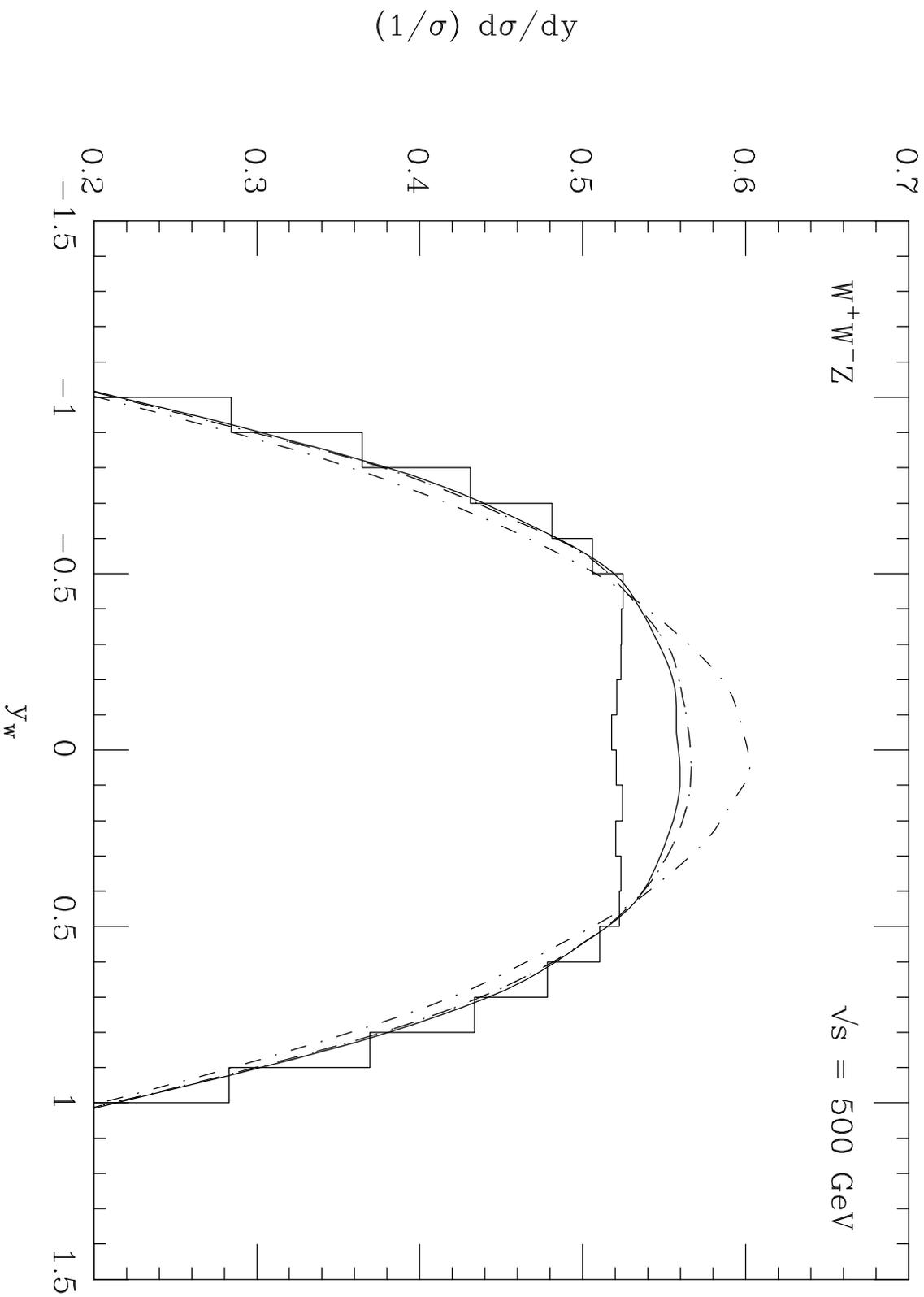
\begin{figure}
\protect
\caption{
Normalized rapidity distribution of the $W^\pm$ bosons  (process
\protect\ref{z}) at $\protect\sqrt{s}= 500$ GeV. The conventions
are the same as in Fig.\ \protect\ref{fig:3}. It is usefull observe
 that the dotted line coincides with  the dashed one.}
\label{fig:4}
\end{figure}

%**********
%Tables
%**********

\begin{table}[htb]
\begin{center}
\begin{tabular}{|c|c|c|c|c|c|c|}
 &\multicolumn{4}{c|}{$WW\gamma$}&\multicolumn{2}{c|}{$WWZ$}\\ \cline{1-5}
 &\multicolumn{2}{c|}{$P_T^\gamma >10$ GeV}&\multicolumn{2}{c|}
{$P_T^\gamma >20$ GeV}&\multicolumn{2}{c|}{}\\ \cline{1-7}
$\protect\sqrt{s}$
&0.5 TeV& 1 TeV&0.5 TeV& 1 TeV&0.5 TeV& 1 TeV
\\ \hline
$\sigma_{\text{sm}}$ ($\times 10^2$ fb)
& 2.55 & 7.83 & 1.45 & 5.21 & 1.90 $\times 10^{-1}$ & 2.38
\\ \hline
$\sigma^0_{\text{int}}$ ($\times 10^2$ fb)
& 1.53 & 9.81 & 9.50 $\times 10^{-1}$ & 7.25 & 2.50 $\times 10^{-1}$
& 6.13
 \\
$\sigma^0_{\text{ano}}$ ($\times 10^2$ fb)
&1.44 $\times 10^{1}$ & 3.25 $\times 10^3$ & 8.57 &2.29 $\times 10^3$
& 1.08 & 1.02 $\times 10^3$
\\ \hline
$\sigma^c_{\text{int}}$ ($\times 10^2$ fb)
& 7.10  $\times 10^{-1}$ &4.65 & 3.20  $\times 10^{-1}$ & 2.68 &
1.00 $\times 10^{-1}$  & 2.74
\\
$\sigma^c_{\text{ano}}$ ($\times 10^2$ fb)
&1.07& 2.33 $\times 10^2$ & 6.00 $\times 10^{-1}$ & 1.63 $\times 10^2$ &
9.00 $\times 10^{-2}$  & 7.61 $\times 10^{1}$
\\ \hline
$\sigma^n_{\text{int}}$ ($\times 10^2$ fb)
&{---}&{---} &{---}&{---}& 0 & 0
\\
$\sigma^n_{\text{ano}}$ ($\times 10^2$ fb)
&{---}&{---}&{---}&{---}& 1.70 $\times 10^{-1}$  & 5.51
\\
\end{tabular}
\end{center}
\caption{Cross sections
$\sigma^i_{\text{sm}}$, $\sigma^i_{\text{int}}$, and
$\sigma^i_{\text{ano}}$, {\bf assuming $\Lambda = M_W$}.}
\label{sigmas:z}
\end{table}

%{\tiny
\begin{table}[htb]
\begin{center}
\begin{tabular}{|c|c|c|c|c|c|c|}
     & \multicolumn{4}{c|}{$WW\gamma$} & \multicolumn{2}{c|}{$WWZ$}\\
\cline{2-5}
 & \multicolumn{2}{c|}{$p_T^\gamma >10$ GeV} & \multicolumn{2}
{c|}{$p_T^\gamma >20$ GeV} & \multicolumn{2}{c|}{ } \\ \hline
$\protect{\sqrt{s}}$
     & $0.5$ TeV & $1$ TeV & $0.5$ TeV & $1$ TeV & $0.5$ TeV & $1$ TeV
\\ \hline
$a_0$ & $(-0.17,0.062)$ & $(-0.011,0.0077)$ & $(-0.18,0.072)$ &
$(-0.011,0.0083)$ & $(-0.34,0.11)$ & $(-0.015,0.0094)$\\
$a_c$ & $(-0.83,0.17)$ & $(-0.045,0.025)$ & $(-0.78,0.24)$ &
$(-0.046,0.029)$ & $(-1.4,0.32)$ & $(-0.065,0.029)$\\
$a_n$ & --- & --- & --- & --- & $(-1.6,1.6)$ & $(-0.16,0.16)$\\
$|\Delta\sigma|$ & $15$ & $26$ & $12$ & $22$ & $4.1$ & $14$ \\
\end{tabular}
\end{center}
\caption{ Allowed intervals of $a_i$ corresponding to an effect smaller
than $3\sigma$ in the total cross section. We also
exhibit the difference ($\Delta\sigma$) between the anomalous
cross sections and the SM ones in fb for a $3\sigma$ effect.}
\label{const}
\end{table}
%}

\end{document}